\documentclass[11pt]{article}
\usepackage{vietnam,epsfig}

\def\be{\begin{equation}}
\def\ee{\end{equation}}
\def\bea{\begin{eqnarray}}
\def\eea{\end{eqnarray}}

\begin{document}
\vspace*{4cm}
\title{DARK ENERGY - A PEDAGOGIC REVIEW}

\author{ PAUL H. FRAMPTON }

\address{Department of Physics and Astronomy, University of North Carolina,\\
Chapel Hill, NC 27599-3255}

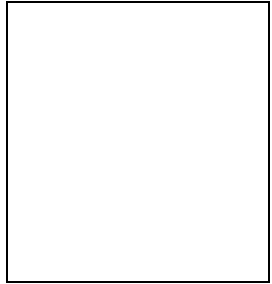
\maketitle\abstracts{
In an introductory manner, the nature of dark energy
is addressed, how it is observed and what further tests
are needed to reconstruct its properties. Several theoretical
approaches to dark energy will be discussed.
}

\section{Plan of the Talk}
\begin{itemize}
\item What observations and theoretical assumptions underly dark energy (DE)?
\item If general relativity (GR) holds at all length scales, the most
conservative assumption, then DE follows from the supernovae Type 1A (SNe1A)
or, independently, from the Cosmic Microwave Background (CMB)
combined with Large Scale Structure (LSS).
\item Should we seriously query GR at large distance scales?

\end{itemize}

\section{Einstein-Friedmann Equation}

The Einstein equations relate geometry on the Left-Hand-Side (LHS)
to the distribution of mass-energy on the Right-Hand-Side (RHS)

\begin{equation}
G_{\mu\nu} = - 8 \pi G T_{\mu\nu}
\label{EinFried}
\end{equation}

We hesitate to change the LHS but it is really checked with precision
only at Solar System (SS) scales. At cosmological length scales, 
we may consider using a modification
such as higher-dimensional gravity.

On the RHS, if we include only luminous and dark matter it is
insufficient (keeping the LHS intact) and there is needed
a further term which could be a cosmological constant or, 
more generally, dark energy.

\section{Observational Issues}

How can we constrain DE?

\begin{itemize}

\item Measurement of the expansion history H(t)
\item The time-dependence of the equation of state w(t)
\item Looking for any clustering property of DE. No evidence for this presently.
\item How does DE couple to Dark Matter (DM)? This
is related to the question of clustering.
\item Local tests of GR and the equivalence principle, though the
extrapolation from the SS to the Universe is
some 13-15 orders of magnitude comparable to the
extrapolation from the
weak scale to the GUT scale in particle phenomenology.
The usual prior is a desert hypothesis.
\end{itemize}

\section{$\Lambda$ as DE: Why $10^{-122}$ (Planck Mass)$^4$?}

We know from the Lamb Shift and Casimir Effect in quantum
electrodynamics that vacuum fluctuations are 
real effects.

If we calculate the value of $\Lambda$, it will naively be
ultra-violet (UV) quartically divergent. The most natural UV cut-off in GR
is the Planck mass $\sim 10^{19} GeV$ whereupon
\begin{equation}
\Lambda \sim (10^{19}GeV)^4 = (10^{28} eV)^4 = 10^{112} (eV)^4
\end{equation}
If we use, instead, the weak scale $\sim 100 GeV$ as our UV cut-off, we arrive
at
\begin{equation}
\Lambda \sim (100 GeV)^4 = (10^{11} eV)^4 = 10^{44} (eV)^4
\end{equation}
The observed value for $\Lambda$, by contrast, is approximately
\begin{equation}
\Lambda \sim (3 \times 10^{-3} eV)^4 \sim  10^{-10} (eV)^4
\end{equation}

\section{Coincidence Problem}

As if the fine-tuning problem for $\Lambda$ were not enough, there
is a second problem with $\Lambda$, the coincidence problem.
Let us define $\Omega_{\Lambda} = \rho_{\Lambda}/\rho_{C}$
as the fraction of the critical density $\rho_{C}$.

The present value is $\Omega_{\Lambda} \sim 0.7$ but it scales,
since $\rho_{\Lambda}$ is constant and assuming $\Omega_{TOT} = 1$,
like $\rho_{C}^{-1} \sim (1 + Z)^{-3}$ so at a redshift $Z > 10$ it was
$\Omega_{\Lambda} < 0.001$ while for a future redshift $Z < -0.9$
one has $\Omega_{\Lambda} > 0.999$. 

If we plot $\Omega_{\Lambda}$ versus log R over cosmic history
from $ - 60 < log_{10} R < + 60$, it appears like a step function
changing from zero to one abruptly around the present era.
Even more dramatic is a plot of $ d\Omega_{\Lambda}/dR$
which approximates a Dirac delta function and the coincidence
problem is then why we live right in the middle of the spike of
the delta function.

If the dark energy had appeared earlier it would have interfered
with structure formation: if later, we would still be unaware
of it.

\section{The Quintessence Possibility}

One parametrization of the dark energy can be made using a 
dynamical scalar field, now generically called quintessence.

\subsection{Scaling potentials}

Examples are:

\begin{equation}
V \sim e^{-\lambda \Phi}
\end{equation}
as in \cite{W,FJ}
\begin{equation}
V \sim ((\Phi - A)^2 + C)e^{ - \lambda \Phi}
\end{equation}
as in \cite{AS}.

\subsection{Tracker Potentials}

Examples are 

\begin{equation}
V \sim \Phi^{- \alpha}
\end{equation}
as in \cite{RP},
\begin{equation}
V \sim exp \left( \frac{M}{Q} - 1 \right)
\end{equation}
as in \cite{SWZ}.

\subsection{Approaches to the Coincidence Problem}

We may assume that our universe sees periodic
epochs of acceleration\cite{DKS} with potential

\begin{equation}
V \sim M^4 e^{- \lambda \Phi} (1 + A \sin a \Phi)
\end{equation}

\bigskip

Another possibility is that it is important
that our epoch is close to the matter/radiation equality
time. This may be incorporated by having a non-minimal
coupling to matter\cite{BM}, to gravity\cite{PB}
or in a k-essence theory with a non-trivial
kinetic term in the lagrangian\cite{AP}.

\bigskip

\section{Dark Energy with Equation of State $w = p/\rho < -1$}

Present data on SNe1A, CMB and LSS are consistent with w=-1 as
for a cosmological constant.

Since the possibility that $w < -1$ is still allowed\cite{MMOT}, 
I shall spend a disproportionate amount of time on it because, if
it persisted, it could well signal new physics.

One interpretation of dark energy comes from string theory, closed strings
on a toroidal cosmology\cite{BFM}. This leads generically to
$w < -1$ \cite{F}.

In general, without dark energy (as in most cosmology texts pre-1998),
the destiny of the Universe was tied to geometry in
a simple manner: the Universe will expand forever
if it is open or flat; it will stop expanding and contract to a Big Crunch if it is closed.

With Dark Energy, this connection between geometry and destiny
is lost and the future fate depends entirely on how the
presently-dominant dark energy will evolve.

This question is studied in \cite{K,FT,RRC}. If $w < -1$ and
is time-independent, the scale
factor diverges at a finite future time
- the Big Rip.
Generally, this will be at least as far in the future as the
Big Bang was in the past.

Such a cosmology may have philosophical appeal? There is
more symmetry between past and future.

If one allows a time-dependent w(t), there are two other 
possible fates:

(i)An infinite-lifetime universe where dark
energy dominates at all future times.

(ii)A disappearing dark energy where the Universe
becomes (again) matter dominated.

The case $w < -1$ gives rise to some exceptionally interesting puzzles for
theoretical physics.

There is the question of violation of the weak energy condition
universally assumed in general relativity. This means
there are inertial frames where the energy density is negative
signaling vacuum instability\cite{F2004,CMT}.

Let us make three assumptions, any or all of which may be incorrect,
just so that we may say something more: that
(i) There is a stable vacuum with $\Lambda = 0$;
(ii) The dark energy decays to it by a 1st-order phase transition;
(iii) There is some, albeit feeble, interaction between dark energy
and the electromagnetic field.

Then one can use old arguments\cite{F1976} to investigate nucleation.
The result is that\cite{F2004} even with the tiniest coupling of dark energy to
the electromagnetic field the dark energy
would have spontaneously decayed long ago unless the appropriate
bubble radius is at least galactic in size. 

In this model, because the energy density of the DE is so small
compared to {\it e.g.} the energy
density in a common macroscopic magnetic field of, say, 10T
the 1st order phase transition can be adequately suppressed
only by decoupling the DE completely from
all but gravitational forces or by
arguing that a collision would need to be between
galaxies or larger objects to be effected.
Certainly, no terrestrial experiment can be influenced:
for one contrary suggestion of a Josephson junction
experiment which might well be justified for other reasons, 
see {\it e.g.} \cite{BeckMackey}.

Of course, this is only a toy model but the general
conclusion is probably correct - that there can
be no microscopic effect of the dark energy.

This makes the DE very difficult or impossible to
investigate except through
astronomical observations.

\section{Dark Energy and Neutrinos}

It has been pointed out by many theorists that
the density of the dark energy $\sim (10^{-3} eV)^4$
is suggestive of the neutrino mass.

Very interesting attempts to strengthen such a 
connection have been made
\cite{PQ,Nelson}. Such 
mass-varying neutrino models
seek to make a direct identification of the DE density
with neutrino mass\cite{F1,F2} itself.

\section{Precision Experiment}

We know well of the precision experiments to test Newton's
Law of Gravity down to a distance of 100 microns and below.

One originator of such ideas suggests\cite{DGZ} a different precision test,
of the Earth-Moon distance, to a similar accuracy of 100 microns, presumably
the distance between the centers of mass. A particular
modification of gravity\cite{DGP} might have a tiny
effect on our lunar system. Clearly if this experiment
can be achieved, the present accuracy being at the level of
centimeters, it would be an impressive achievement.

\section{Conclusions: Observation and Theory}

\begin{itemize}

\item The theoretical community has yet to come up with a definitive proposal
to explain the dark energy.

\item The nature of the dark energy is so profound for cosmology
and particle physics that we desperately need more SNe1A observations from
important proposed experiments {\it e.g.} SNAP (for which NASA funding has sadly been
suspended for 5 years as a result of prioritizing sending humans to Mars!), as well
as complementary observational constraints on the CMB from {\it e.g.} the Planck mission.

\item The equation of state will be decisive. If w=-1, it's a cosmological
constant with its fine-tuning and coincidence problems. If $w > -1$
quintessence will receive a shot in the arm.

\item If the data would settle down to a value $w < -1$ we could be at the dawn
of a revolution in theory with general relativity at the largest distance scales
called into question.

\end{itemize}

\section*{Acknowledgments}
We thank Tran Thanh Van for the invitation to Hanoi.
This work was supported in part by the US Department of Energy
under Grant No. DE-FG02-97ER-41036.

\end{document}